\documentclass[conference]{IEEEtran}
\IEEEoverridecommandlockouts

\usepackage{cite}
\usepackage{amsmath,amssymb,amsfonts}
\usepackage{algorithmic}
\usepackage{graphicx}
\usepackage{textcomp}
\usepackage{xcolor}
\usepackage{multirow}
\usepackage{subfig}
\usepackage{booktabs}
\usepackage{url}

\begin{document}

% -------------------- Title --------------------
\title{\fontsize{20}{22}\selectfont \textbf{SHIELD8-UAV:} Sequential 8-bit Hardware Implementation\\ of a Precision-Aware 1D-F-CNN for Low-Energy\\ UAV Acoustic Detection and Temporal Tracking}

% -------------------- Authors --------------------
\author{
\IEEEauthorblockN{Susmita Ghanta}
\IEEEauthorblockA{\textit{Dept. of Electrical Engineering}\\
\textit{Indian Institute of Technology Jammu}\\
Jagti 181221, India\\
2024pvl0105@iitjammu.ac.in}
 \and
 \IEEEauthorblockN{Karan Nathwani}
 \IEEEauthorblockA{\textit{Dept. of Electrical Engineering}\\
 \textit{Indian Institute of Technology Jammu}\\
 Jagti 181221, India\\
 karan.nathwani@iitjammu.ac.in}
 \and
 \IEEEauthorblockN{Rohit Chaurasiya}
 \IEEEauthorblockA{\textit{Dept. of Electrical Engineering}\\
 \textit{Indian Institute of Technology Jammu}\\
 Jagti 181221, India\\
rohit.chaurasiya@iitjammu.ac.in}
}

\maketitle

\begin{abstract}
Real-time unmanned aerial vehicle (UAV) acoustic detection at the edge demands low-latency inference under strict power and hardware limits. This paper presents SHIELD8-UAV, a sequential 8-bit hardware implementation of a precision-aware 1D feature-driven CNN (1D-F-CNN) accelerator for continuous acoustic monitoring. The design performs layer-wise execution on a shared multi-precision datapath, eliminating the need for replicated processing elements. A layer-sensitivity quantisation framework supports FP32, BF16, INT8, and FXP8 modes, while structured channel pruning reduces the flattened feature dimension from 35,072 to 8,704 (75\%), thereby lowering serialised dense-layer cycles. The model achieves 89.91\% detection accuracy in FP32 with less than 2.5\% degradation in 8-bit modes. 

The accelerator uses 2,268 LUTs and 0.94 W power with 116 ms end-to-end latency, achieving 37.8\% and 49.6\% latency reduction compared with QuantMAC \cite{QuantMAC} and LPRE \cite{LPRE}, respectively, on a Pynq-Z2 FPGA, and 5–9\% lower logic usage than parallel designs. ASIC synthesis in UMC 40 nm technology shows a maximum operating frequency of 1.56 GHz, 3.29 mm\textsuperscript{2} core area, and 1.65 W total power. These results demonstrate that sequential execution combined with precision-aware quantisation and serialisation-aware pruning enables practical low-energy edge inference without relying on massive parallelism.
\end{abstract}

% -------------------- Keywords --------------------

\section{Introduction}
The rapid growth of edge artificial intelligence (AI) has intensified the demand for energy-efficient neural network accelerators capable of real-time inference under constrained power and hardware budgets. Among emerging edge workloads, unmanned aerial vehicle (UAV) acoustic detection has gained significant attention for surveillance, security, and autonomous monitoring applications, where reliable operation is required under low-visibility and non-line-of-sight conditions. Conventional CNN accelerators primarily rely on spatial parallelism through replicated processing elements to achieve high throughput. While effective in data-centre environments, such architectures incur substantial area, memory bandwidth, and power overhead, making them unsuitable for edge deployments. Moreover, dense layers and high-precision computation further increase latency and energy consumption, limiting scalability on resource-constrained platforms.

\begin{figure}[!t]
    \centering
    \includegraphics[width=0.875\columnwidth]{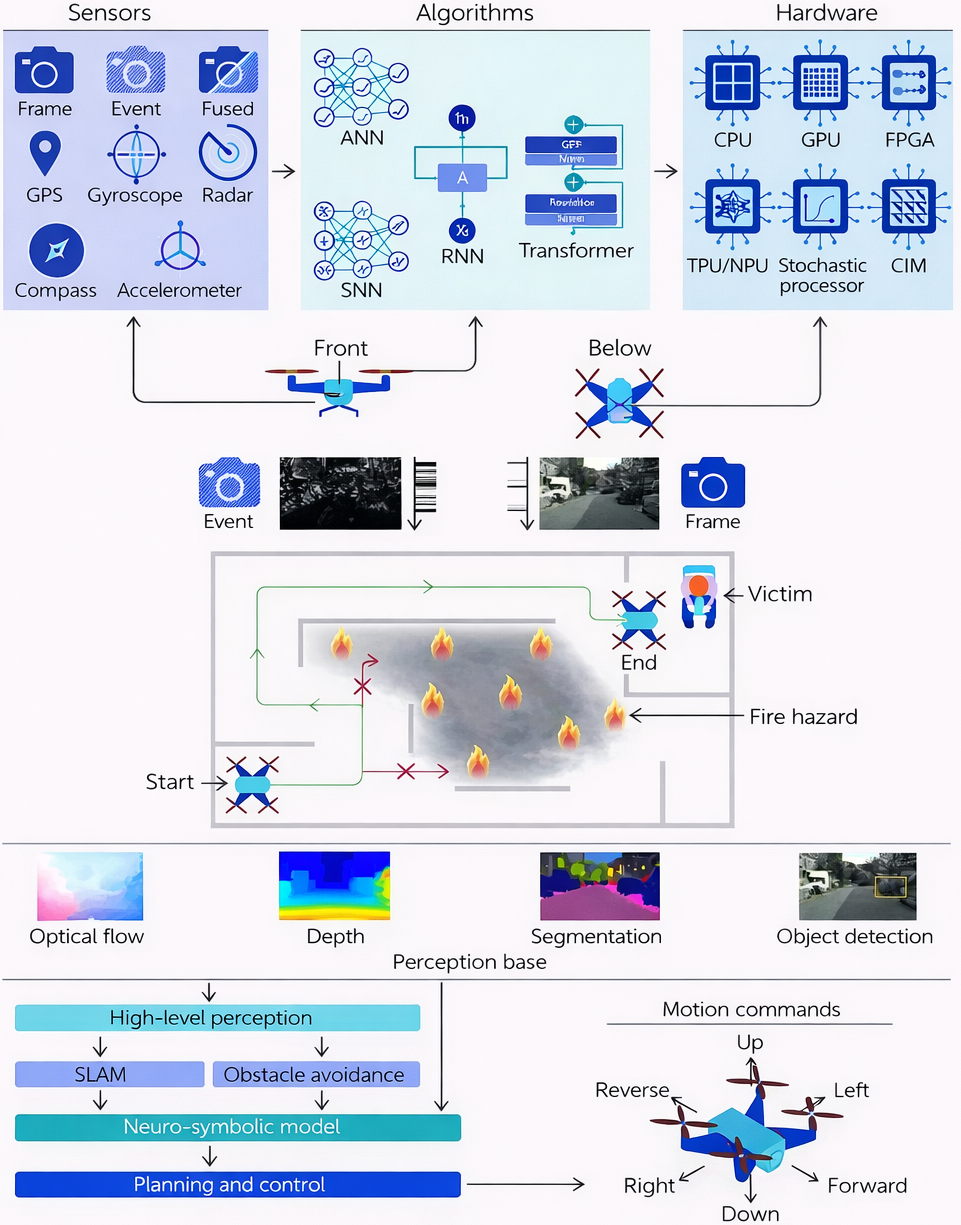}
    \caption{Overview of the algorithm-hardware co-design stack for UAV acoustic detection and temporal tracking on edge platforms.}
    \label{fig:overview_stack}
\end{figure}

Recent research has explored model compression and low-precision inference to improve hardware efficiency; however, many existing designs still depend on partially parallel compute fabrics and lack unified hardware–algorithm co-optimisation. In particular, redundant datapath duplication and unstructured computation patterns lead to inefficient utilization of on-chip resources during sequential layer execution. Fig.~\ref{fig:overview_stack} illustrates the proposed co-design stack spanning sensing, feature extraction, learning, and hardware mapping. Acoustic signals are converted into compact feature vectors, processed by the lightweight 1D-F-CNN, and executed on a sequential multi-precision accelerator with shared datapath reuse. The figure highlights the system-level integration between algorithm design and hardware execution, which forms the central contribution of SHIELD8-UAV.

To address these limitations, this work proposes a hardware-aware sequential CNN accelerator tailored for edge UAV acoustic detection. The design employs a shared reconfigurable datapath that executes convolutional and fully connected layers in a serialised manner, eliminating the need for hardware replication while maintaining functional flexibility. In addition, a precision-aware quantisation strategy enables adaptive computation across multiple numeric formats, and structured pruning is incorporated to reduce dense-layer serialisation overhead. The primary contributions of this work are as follows:

\begin{itemize}

\item A reusable sequential layer-execution CNN accelerator that eliminates datapath replication by mapping convolutional and dense layers onto a shared compute fabric, reducing FPGA logic usage to 2,268 LUTs, approximately 5–9$\times$ smaller than representative parallel CNN accelerators.

\item A layer-sensitivity-driven multi-precision inference framework supporting FP32, BF16, INT8, and FXP8 formats, achieving 89.91\% detection accuracy with less than 2.5\% degradation in 8-bit operation.

\item A serialisation-aware structured channel pruning strategy that reduces the flattened feature dimension from 35,072 to 8,704 (75\% reduction), directly decreasing dense-layer execution cycles and verification complexity.

\item The algorithm–hardware co-design validated on FPGA and 40 nm ASIC, achieving 116 ms real-time inference latency at 0.94 W and outperforming prior reusable edge accelerators, including QuantMAC\cite{QuantMAC} and LPRE\cite{LPRE} by 37.8\% and 49.6\% latency reduction, respectively.

\end{itemize}

Experimental results demonstrate that the proposed co-designed framework achieves efficient real-time inference with reduced computational complexity and hardware utilization, making it well-suited for low-power edge-AI systems.

\section{Background and Motivation}

\subsection{Acoustic UAV Detection and Feature-Driven Learning}
UAV detection has been explored using RF, radar, vision, and acoustic sensing modalities. Among these, acoustic sensing is well-suited for edge deployment due to its low sensor cost, passive operation, and robustness under non-line-of-sight and low-illumination conditions. Early approaches relied on handcrafted acoustic features with shallow classifiers, whereas recent methods employ deep learning models on time–frequency representations for improved robustness.

Wavelet scattering transform integrated with 1D-CNN architectures has shown strong performance in UAV acoustic classification \cite{WST_CNN}, and large-scale audio-based CNN models further validate the effectiveness of learned feature representations \cite{audio-based}. More generally, 1D-CNNs are suitable for environmental sound analysis as they capture temporal dependencies while maintaining a compact parameter footprint \cite{environmental}. However, continuous edge monitoring imposes strict power, memory, and latency constraints. In such settings, repeated feature-map transfers and dense-layer serialisation can dominate execution cost, motivating lightweight feature-driven networks that balance accuracy and hardware efficiency.

\subsection{Sequential, Precision-Aware Edge Hardware}
Edge inference workloads are dominated by MAC computation and memory movement, requiring architectures optimized for compute efficiency and bandwidth use. Conventional accelerators employ spatial parallelism with replicated processing elements, leading to high area, routing, and power overheads on resource-constrained platforms. A sequential layer-reuse approach instead schedules all layers on a shared compute fabric, improving hardware utilization while avoiding datapath duplication. Efficiency is further enhanced through precision scaling, where reduced operand bit width lowers bandwidth demand and switching activity. Prior studies show that precision-aware quantisation and multi-precision MAC units can retain model accuracy while improving energy efficiency \cite{POLARON}, and recent accelerator trends emphasise algorithm–hardware co-design and utilisation-oriented optimisation rather than peak throughput \cite{Norrie2022}.

Despite these advances, dense-layer input serialization remains a major bottleneck in sequential edge accelerators, contributing to both latency and verification overhead. This work therefore pursues two objectives: (i) enabling reusable sequential inference using a unified multi-precision compute datapath that overlaps activation latency with MAC data loading, and (ii) co-designing quantisation and pruning to directly reduce serialisation overhead and scheduling cost for continuous edge monitoring workloads.

\begin{figure}[!t]
    \centering
    \includegraphics[width=0.925\columnwidth]{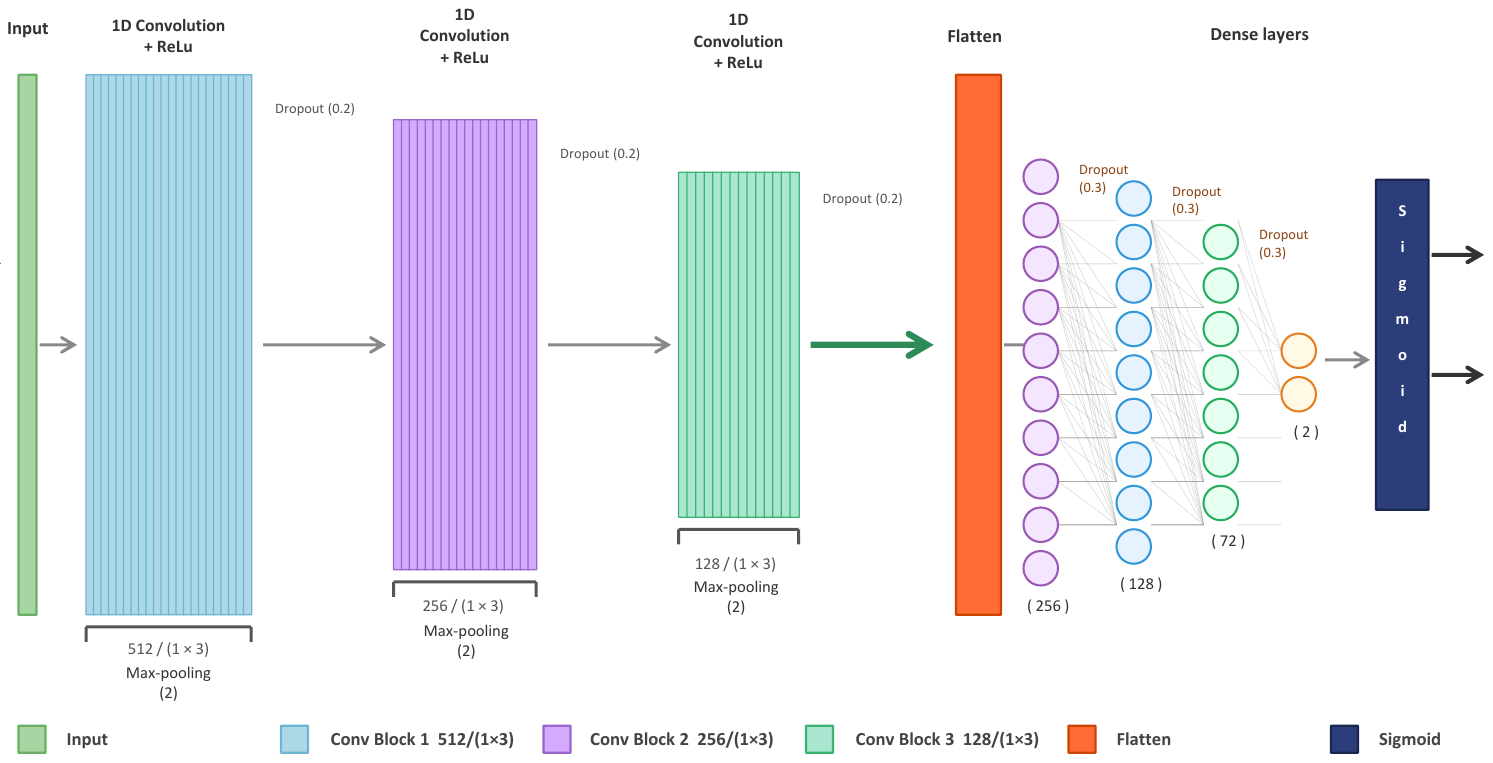}
    \caption{The proposed 1D-F-CNN architecture used for acoustic UAV detection. The network operates on compact feature sequences and is designed for efficient mapping to sequential hardware execution.}
    \label{fig:fcnn_model}
\end{figure}

\section{Proposed SHIELD8-UAV Approach}

SHIELD8-UAV adopts an algorithm—hardware co-design framework in which the network architecture, quantisation scheme, and accelerator design are jointly optimised for sequential edge inference. The proposed system integrates four key components: (i) a compact feature-driven 1D-F-CNN, (ii) layer-wise precision-aware quantisation, (iii) structured pruning for serialisation-aware optimisation, and (iv) a reusable multi-precision sequential accelerator architecture.

\subsection{Feature-Driven 1D-F-CNN for UAV Detection}
The proposed model operates on compact acoustic feature vectors extracted from short-duration audio segments, avoiding computationally expensive 2D spectrogram processing. This design reduces memory footprint, feature-map storage, and data movement, enabling efficient hardware mapping.

Given an input feature vector $\mathbf{x}\in \mathbb{R}^{1\times M}$, each convolutional block performs 1D convolution followed by activation, pooling, and regularisation:
\begin{equation}
\mathbf{o}=\mathcal{D}_{0.2}\Big(\mathcal{M}_{1\times2}\big(\sigma_R(\mathcal{C}_{1\times3}(\mathbf{x}))\big)\Big),
\end{equation}
where $\mathcal{C}_{1\times3}$ denotes the 1D convolution, $\sigma_R(\cdot)$ is ReLU, $\mathcal{M}_{1\times2}$ represents max-pooling, and $\mathcal{D}_{0.2}$ denotes dropout. Three convolutional stages capture temporal acoustic patterns such as rotor harmonics and periodic signatures, followed by dense layers for binary classification. Compared to 2D CNNs, the 1D architecture significantly reduces parameter count and intermediate feature size while preserving discriminative temporal representation \cite{WST_CNN}.

\subsection{Precision-Aware Quantisation With Layer Sensitivity}
To balance accuracy and hardware efficiency, a layer-wise precision assignment strategy is employed based on quantisation sensitivity. For each layer $l$, a sensitivity score is defined as
\begin{equation}
s_{l,\mathrm{sc},k} =
\frac{\left(\left\|Q^{\mathrm{PwQ}}(\mathbf{w}_l)-\mathbf{w}_l\right\| -
\left\|Q^{\mathrm{PwQ}}_{\mathrm{sc},k}(\mathbf{w}_l)-\mathbf{w}_l\right\|\right)
\cdot \left\|\nabla \mathcal{L}_{\mathbf{w}_l}\right\|}{n_l},
\end{equation}
\begin{equation}
s_l = \max\left(s_{l,\mathrm{sc},16},\, s_{l,\mathrm{sc},8}\right),
\end{equation}
where $\mathbf{w}_l$ and $n_l$ denote the weights and size of layer $l$, respectively. Layers with higher sensitivity are assigned higher precision (FP32/BF16), while less sensitive layers operate in INT8 or FXP8 to reduce arithmetic cost.

Weights are quantised using learned clipping bounds $W_l$ and $W_h$. For a full-precision weight $W$ and bit-width $n$, the scale factor is
\begin{equation}
\mathrm{scale}(k) = \mathrm{mean}(|W|)\cdot \frac{2^n-1}{2^{n-1}},
\end{equation}
and the integer quantised weight is
\begin{equation}
\widehat{W} =
\mathrm{round}\!\left(
\left(\mathrm{clip}\!\left(\frac{W}{k}, W_l, W_h\right)-W_l\right)
\cdot \frac{2^n-1}{W_h-W_l}
\right),
\end{equation}
with reconstruction
\begin{equation}
Q^{\mathrm{PwQ}}(W) = \widehat{W}\cdot \frac{W_h-W_l}{2^n-1}+W_l.
\end{equation}

Activation quantisation follows the PACT formulation with a learnable clipping parameter $\alpha$:
\begin{equation}
y = \mathrm{PACT}(x) = 0.5\left(|x| - |x-\alpha| + \alpha\right),
\end{equation}
\begin{equation}
x^q =
\mathrm{round}\!\left(y\cdot \frac{2^n-1}{\alpha}\right)
\cdot \frac{\alpha}{2^n-1}.
\end{equation}
This formulation enables flexible multi-precision inference across FP32, BF16, INT8, and FXP8 modes while preserving detection accuracy compared to \cite{QuantMAC}.

\subsection{Structured Pruning for Serialisation-Aware Optimisation}
In sequential accelerators, the flatten-to-dense interface becomes a dominant latency bottleneck due to serialised data movement through a shared datapath. To address this, structured channel pruning is applied before flattening, reducing the feature dimension from 35,072 to 8,704. This directly decreases dense-layer MAC operations and execution cycles in serialized hardware pipelines. Unlike conventional compression-oriented pruning, the objective here is scheduling-aware optimisation that improves latency, simulation efficiency, and verification feasibility in reusable accelerator designs.  Table~\ref{tab:pruning} discusses the implications in detail. 

\begin{table}[!t]
\centering
\caption{Dense-Layer Feature Reduction and Hardware Benefits}
\label{tab:pruning}
\begin{tabular}{|l|c|c|}
\hline
\textbf{Metric} & \textbf{Before Pruning\cite{WST_CNN}} & \textbf{After Pruning} \\
\hline
Flatten size & 35,072 & 8,704 \\
\hline
Size reduction & \multicolumn{2}{c|}{75\%} \\
\hline
Dense MACs & Very high & $\approx$75\% lower \\
\hline
Serialized cycles & 35,072 & 8,704 \\
\hline
Design debug & High & Reduced \\
\hline
Timing \& power & Challenging & Improved \\
\hline
\end{tabular}
\end{table}

\subsection{Sequential Multi-Precision Accelerator Architecture}
\begin{figure}[t]
    \centering
    \includegraphics[width=0.925\columnwidth]{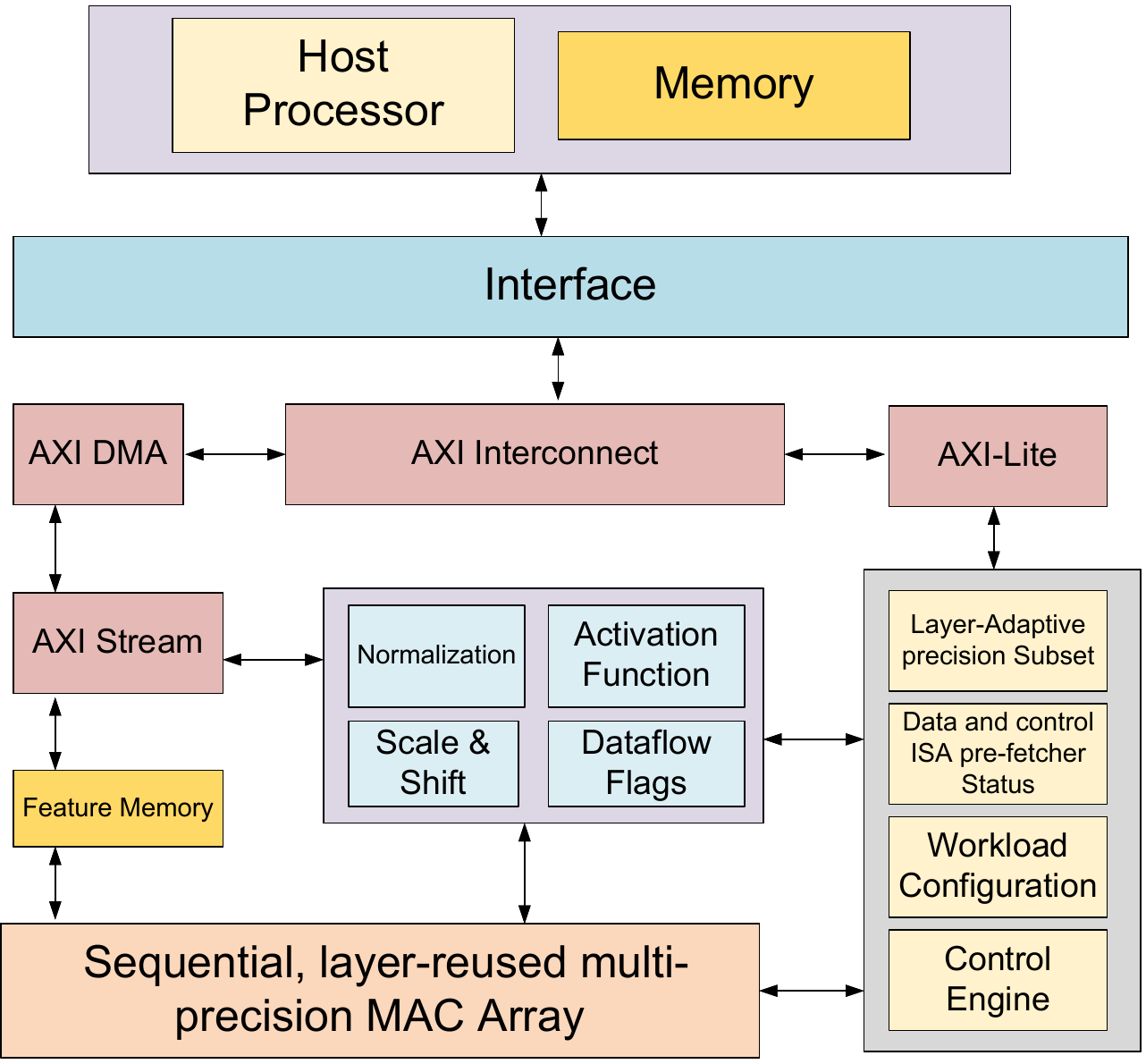}
    \caption{Proposed precision-aware sequential accelerator architecture with shared compute and multi-precision support.}
    \label{fig:arch}
\end{figure}

The proposed POLARON accelerator adopts a reusable shared-compute datapath composed of input and weight buffers, a configurable MAC bank, accumulation and activation units, and an FSM-based control engine (Fig.~\ref{fig:arch}). All neural network layers execute sequentially on the same compute fabric, avoiding dedicated per-layer processing elements and thereby reducing area and control complexity. Feature maps and weights are streamed from on-chip buffers to the MAC bank, and intermediate activations are written back to local memory for reuse in subsequent layers. The datapath supports FP32, BF16, INT8, and FXP8 computations via programmable alignment and scaling logic, while extended-precision accumulators maintain numerical stability during reduced-precision operations. The controller orchestrates convolution, pooling, and fully connected operations within a unified execution pipeline for end-to-end network inference.

Communication with the host system is via an AXI interface, with AXI-Lite for configuration and AXI-DMA/Stream for data transfer. Incoming data is staged in on-chip feature memory prior to computation. Per-layer precision selection allows adaptation of numeric format based on workload requirements to balance energy and accuracy. After accumulation, results pass through normalization, scale-and-shift, and overflow handling before entering a CORDIC-based activation unit supporting Swish, SoftMax, SeLU, GELU, Sigmoid, Tanh, and ReLU. A configuration prefetcher interprets layer metadata and updates execution parameters at runtime. The resulting dataflow minimizes memory traffic and latency while enabling precision-aware inference for edge-AI workloads.

\section{Methodology and Experimental Setup}

\subsection{Dataset Preparation and Augmentation}
A heterogeneous acoustic dataset comprising UAV and non-UAV environmental sounds was curated to emulate realistic deployment conditions. UAV recordings were collected from multiple commercial quadrotor platforms at different distances and orientations to capture rotor harmonics, startup transients, and flight-state variations. Background samples include environmental and aircraft-related audio representative of open-field and airport scenarios. 

To improve robustness, recordings were augmented with additive Gaussian noise across a controlled SNR range. Each audio stream was segmented into 0.8-second windows to balance temporal feature capture and inference latency, followed by amplitude normalisation to ensure consistent scaling across samples. Feature extraction was performed using \texttt{librosa}, including MFCC, pooled mel-spectrogram coefficients, power spectral density (PSD), and zero-crossing rate (ZCR), enabling representation of both periodic rotor signatures and broadband acoustic activity. Public audio repositories (e.g., AudioSet and Pixabay) were additionally incorporated to enhance background diversity and reduce environment-specific bias \cite{WST_CNN}.

\subsection{Training and Evaluation Metrics}
The dataset was partitioned into training, validation, and test sets for unbiased evaluation. Model training employed the Adam optimiser with cross-entropy loss and early stopping based on validation performance. Detection performance is reported using accuracy, precision, recall, and F1-score. Given the requirements of continuous monitoring systems, false alarm rate and missed detection rate are also evaluated to assess practical reliability. Robustness was further analysed by measuring inference accuracy across varying SNR levels, reflecting realistic noisy deployment scenarios.

\subsection{Algorithm-Hardware Co-Design Evaluation}
The proposed framework was validated through both algorithm-level emulation and hardware implementation. A Python-based arithmetic emulation model was developed to analyse inference behavior under FP32, BF16, INT8, and FXP8 precision formats prior to RTL realisation. The accelerator was implemented in Verilog HDL and verified using cycle-accurate simulation. FPGA synthesis was performed using the AMD Vivado toolchain on the Pynq-Z2 platform, while ASIC synthesis was conducted using Cadence Genus targeting a UMC 40\,nm CMOS technology library under uniform timing constraints. For fair comparison, representative baseline accelerators were evaluated under normalised assumptions where required. Hardware metrics include LUT/DSP utilisation, operating frequency, power, and estimated area, while system-level evaluation considers end-to-end inference latency and energy consumption during continuous operation.

\begin{table}[t]
\centering
\caption{Evaluation Metrics for UAV Detection Models}
\label{tab:uav_eval}
\renewcommand{\arraystretch}{1.15}
\resizebox{\linewidth}{!}{%
\begin{tabular}{|l|l|c|c|c|c|}
\hline
\textbf{Model} & \textbf{Features} & \textbf{Acc. (\%)} & \textbf{Prec. (\%)} & \textbf{Rec. (\%)} & \textbf{F1 (\%)} \\
\hline
\multirow{3}{*}{SVM\cite{publicsafety}} 
 & Mel (128)          & 87.88 & 88.41 & 87.44 & 87.92 \\
 & MFCC (20)          & 84.39 & 85.62 & 83.17 & 84.37 \\
 & $\log_{10}$(PSD)   & 83.10 & 80.23 & 88.43 & 84.13 \\
\hline
1D-R-CNN\cite{WST_CNN} & Raw audio & 88.90 & 92.78 & 84.54 & 88.47 \\
\hline
2D-CNN\cite{Iot2} & MFCC (20) & 88.96 & 88.85 & 89.32 & 89.09 \\
\hline
KNN\cite{publicsafety} & MFCC (20) & 88.73 & 87.87 & 90.21 & 89.02 \\
\hline
Ensemble\cite{9774975} & Mel (128) & 86.50 & 86.88 & 85.55 & 86.21 \\
\hline
1D-M-CNN\cite{Iot1} & MFCC (128) & 83.26 & 86.11 & 85.52 & 85.81 \\
\hline
\multirow{4}{*}{\textbf{Baseline (FP32)}} 
 & Mel (128)          & 89.13 & 91.64 & 86.48 & 88.99 \\
 & MFCC (20)          & 89.91 & 89.00 & 91.39 & 90.18 \\
 & $\log_{10}$(PSD)   & 87.87 & 88.88 & 86.92 & 87.89 \\
 & ZCR                & 60.64 & 58.61 & 77.58 & 66.77 \\
\hline
\multirow{4}{*}{\textbf{Proposed (BF16)}} 
 & Mel (128)          & 88.94 & 91.42 & 86.21 & 88.71 \\
 & MFCC (20)          & 89.80 & 88.89 & 91.28 & 90.07 \\
 & $\log_{10}$(PSD)   & 87.65 & 88.65 & 86.68 & 87.66 \\
 & ZCR                & 60.14 & 58.11 & 77.08 & 66.27 \\
\hline
\multirow{4}{*}{\textbf{Proposed (INT8)}} 
 & Mel (128)          & 88.02 & 90.31 & 85.06 & 87.73 \\
 & MFCC (20)          & 89.14 & 88.23 & 90.62 & 89.41 \\
 & $\log_{10}$(PSD)   & 86.55 & 87.56 & 85.60 & 86.57 \\
 & ZCR                & 57.34 & 55.31 & 74.28 & 63.47 \\
\hline
\multirow{4}{*}{\textbf{Proposed (FXP8)}} 
 & Mel (128)          & 87.65 & 89.88 & 84.72 & 87.02 \\
 & MFCC (20)          & 88.97 & 88.06 & 90.45 & 89.24 \\
 & $\log_{10}$(PSD)   & 86.02 & 87.03 & 85.07 & 86.04 \\
 & ZCR                & 55.40 & 53.37 & 72.34 & 61.53 \\
\hline
\end{tabular}}
\end{table}

\section{Results and Discussion}

\subsection{Detection Performance and Precision Impact}

Table~\ref{tab:uav_eval} summarises the detection performance of the proposed 1D-F-CNN against classical and deep-learning baselines. Using MFCC features in FP32 mode, the model achieves 89.91\% accuracy and 90.18\% F1-score, demonstrating competitive performance while operating on a compact feature representation. BF16 inference exhibits near-identical accuracy to FP32, indicating low sensitivity to mantissa reduction. INT8 and FXP8 show moderate degradation but remain suitable for practical monitoring scenarios. These results validate the layer-wise precision-aware quantisation strategy in maintaining accuracy under reduced numerical precision.

\begin{figure}[!t]
    \centering
    \includegraphics[width=0.875\columnwidth]{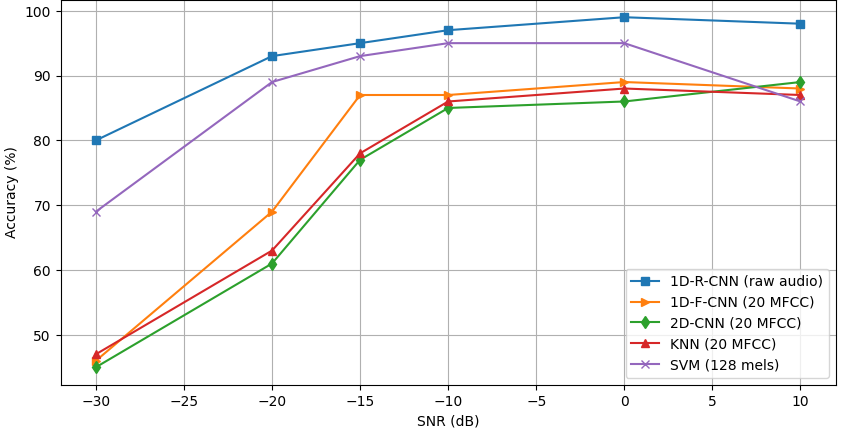}
    \caption{Detection accuracy versus signal-to-noise ratio (SNR).}
    \label{fig:snr_acc}
\end{figure}

\begin{figure}[!t]
    \centering
    \subfloat[False alarm rate]{\includegraphics[width=0.48\columnwidth]{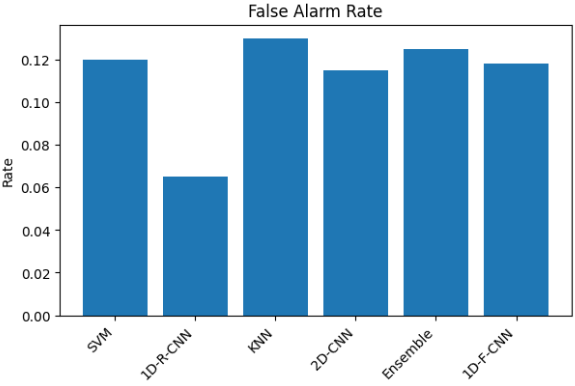}\label{fig:false_alarm}}
    \hfill
    \subfloat[Missed detection rate]{\includegraphics[width=0.48\columnwidth]{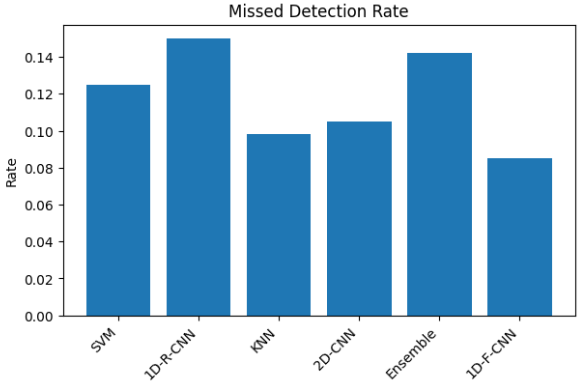}\label{fig:missed_detect}}
    \caption{Error analysis under noise conditions: (a) false alarm rate and (b) missed detection rate. False alarms remain at low SNR levels, while missed detections increase when UAV acoustic signatures are impacted by background noise.}
    \label{fig:error_rates}
\end{figure}

Fig.~\ref{fig:snr_acc} presents accuracy trends across varying SNR levels, and Fig.~\ref{fig:error_rates} shows false alarm and missed detection behavior. The model maintains stable performance under moderate noise, reflecting effective capture of rotor acoustic patterns. False alarms remain low across most SNR ranges, while missed detections increase only at very low SNR due to masking of harmonic signatures. Overall, the feature-driven 1D architecture demonstrates robustness for continuous real-world monitoring. Table~\ref{tab:pruning} shows structured pruning reduces the flattened feature size from 35,072 to 8,704 (75\%), thereby proportionally lowering the number of dense-layer cycles in the sequential shared datapath.
Thus, pruning improves not only model compactness but also inference latency and hardware verification efficiency.

\subsection{Hardware Utilization}

Table~\ref{tab:fpga_resource} compares the proposed architecture with representative FPGA accelerator styles. Fully parallel designs achieve high throughput at the cost of significant LUT, register, and DSP usage. In contrast, the proposed sequential architecture employs a unified compute datapath, minimising resource duplication while preserving functional flexibility. As a result, the design achieves lower hardware utilisation than parallel or partially reusable accelerators. Table~\ref{tab:hydra_compare} further confirms that competitive performance and lower resource overhead through compute reuse and precision adaptation.

ASIC synthesis results at 40~nm (Table~\ref{tab:asic_eval}) indicate a balanced trade-off between frequency, area, and power. Unlike massively parallel accelerators that prioritise peak throughput, the proposed design targets sustained inference efficiency suitable for continuous sensing workloads. The shared multi-precision datapath reduces switching activity and enables scalable deployment across different energy budgets.

\subsection{System-Level Latency and Energy}

A cycle-accurate timing model was developed to analyse latency for reusable and parallel accelerator architectures, as defined in Eqs.~\ref{eq:timing} and~\ref{eq:total_timing}. The complete system was deployed on the Pynq-Z2 platform for end-to-end evaluation. The proposed accelerator achieves 116~ms inference latency under constrained power ($\leq$5W), outperforming prior reusable and edge platforms, including Flex-PE\cite{Flex-PE} (186.4 ms), GR-ACMTr\cite{GR-ACMTr} (772 ms), LPRE\cite{LPRE} (184 ms), QuantMAC\cite{QuantMAC} (163.7 ms), Jetson Nano (226 ms), and Raspberry Pi (555 ms).

\begin{table}[!t]
\caption{FPGA resource utilisation comparison with representative architectural baselines}
\label{tab:fpga_resource}
\renewcommand{\arraystretch}{1.25}
\centering
\resizebox{\linewidth}{!}{%
\begin{tabular}{|l|c|c|c|c|}
\hline
\textbf{Architecture} & \textbf{LUTs} & \textbf{Reg./FFs} & \textbf{BRAM/DSPs} & \textbf{Power (W)} \\
\hline
Fully-parallel \cite{GR-ACMTr} & 20790 & 30684 & 53 & 2.2 \\
Hardware-reused \cite{QuantMAC} & 14428 & 15582 & 23 & 1.28 \\
Layer-reused \cite{Retro} & 13956 & 16323 & 24 & 1.24 \\
Layer-multiplexed \cite{GR-Neuro} & 11265 & 11348 & 32 & 0.73 \\
\textbf{Proposed (SHIELD8-UAV)} & 2,268 & 3250 & 8 & 0.94 \\
\hline
\end{tabular}}
\end{table}

\begin{table}[!t]
\caption{Quantitative FPGA comparison with prior architectural works}
\label{tab:hydra_compare}
\renewcommand{\arraystretch}{1.45}
\centering
\resizebox{\linewidth}{!}{%
\begin{tabular}{|l|c|c|c|c|c|}
\hline
\textbf{Parameters} & \textbf{Lu et al. \cite{9857602}} & \textbf{Aimar et al. \cite{NULLHOP}} & \textbf{Mian et al. \cite{ESL2}} & \textbf{RAMAN \cite{RAMAN}} & \textbf{Proposed} \\
\hline
Platform & Zynq-7100 & VC707 & ZCU102 & Efinix Ti60 & VC707 \\
Model & CNN & CNN & 1D-CNN & DS-CNN & 1D-F-CNN \\
\hline
LUTs (K) & 22.9 & 23.9 & 39 & 37.2 & 2.2 \\
Reg./FFs (K) & 10.7 & 20.1 & 27.8 & 8.6 & 3.25 \\
\hline
Power (W) & 1.1 & 2.2 & 1.54 & 0.15 & 0.94 \\
Freq. (MHz) & 60 & 170 & 200 & 75 & 100 \\
\hline
\end{tabular}}
\end{table}

\begin{equation} \begin{aligned} T_P &= T_{\mathrm{MAC}} + T_{\mathrm{AF}}, \\ T_R &= T_{\mathrm{MAC}} + T_{\mathrm{Serial}} + K \cdot T_{\mathrm{AF}}, \\ &\text{where } T_{\text{MAC}},\ T_{\text{AF}},\ T_{\text{PISO}} \text{is delay for MAC, and serial AF.} \end{aligned} \label{eq:timing} \end{equation} \begin{equation} \begin{IEEEeqnarraybox}[\relax][c]{l} Total\_T_P = \sum_{l=1}^{L-1} n(l) + L - 1, \\ Total\_T_R = \sum_{l=1}^{L} n(l) + 2L - 3, \\ \text{where $L$ and $n(l)$: no. of layers and MACs in layer.} \end{IEEEeqnarraybox} \label{eq:total_timing} \end{equation}

\begin{figure}[!t]
    \centering
    \includegraphics[width=\columnwidth, height=45mm]{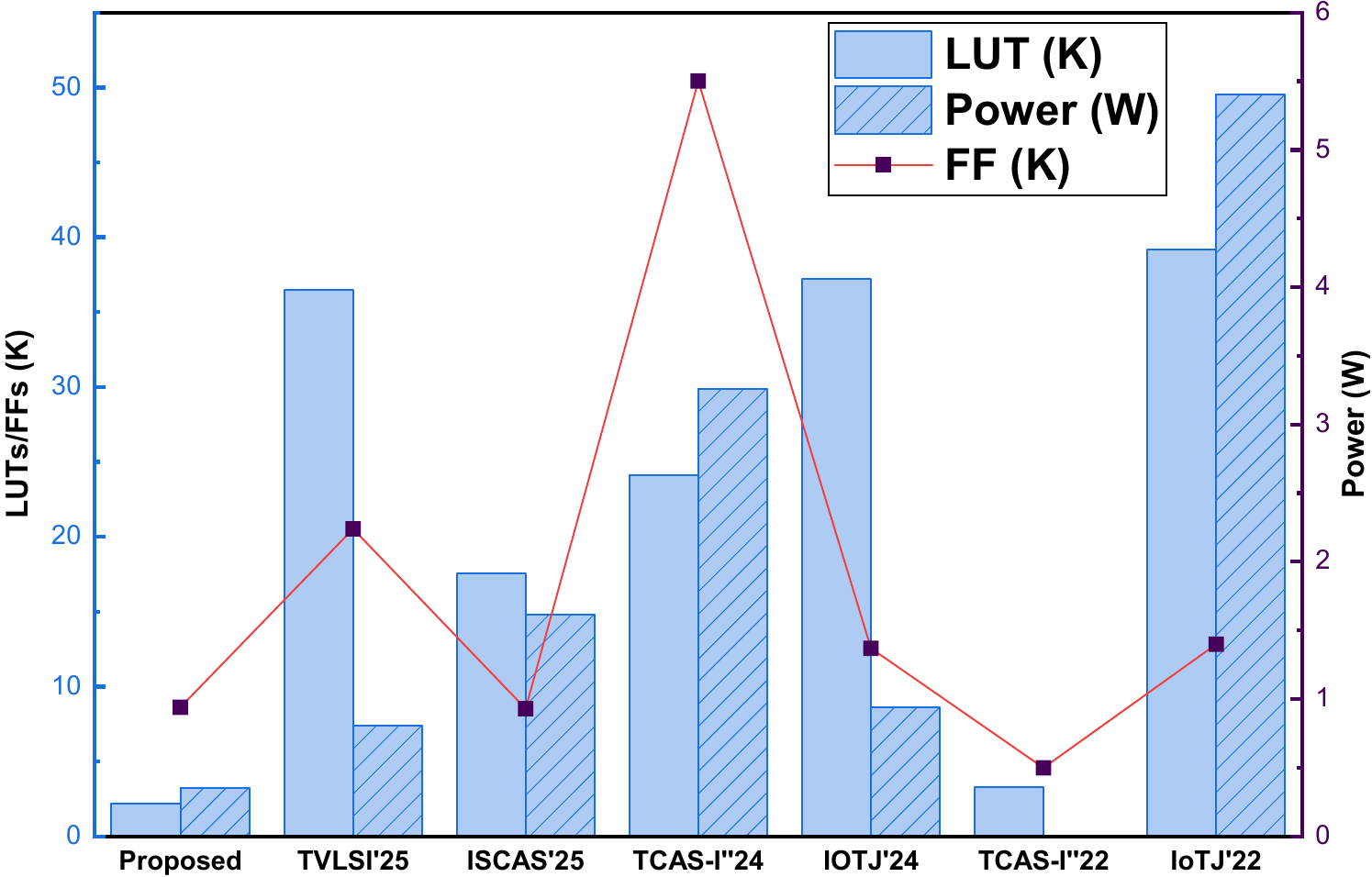}
    \caption{System-level performance summary.}
    \label{fig:system_summary}
\end{figure}

Compared to prior reusable accelerators, the proposed system achieves lower latency and competitive energy efficiency. The improvement is primarily attributed to reduced dense-layer serialisation, precision-aware arithmetic, and unified compute reuse, which collectively minimise memory transfers and execution overhead. Fig.~\ref{fig:system_summary} summarises the overall system performance, demonstrating that sequential execution combined with hardware-aware co-design enables practical real-time edge inference.

\section{Conclusion \& Future Work}

This paper presented SHIELD8-UAV, a precision-aware sequential 1D-F-CNN accelerator for continuous UAV acoustic monitoring on resource-constrained edge platforms. By combining reusable layer-wise execution, multi-precision computation, and serialisation-aware optimisation, the proposed system achieves 89.91\% detection accuracy with less than 2.5\% degradation under 8-bit inference (INT8/FXP8). Structured channel pruning reduces the flattened feature dimension from 35,072 to 8,704 (75\% reduction), directly lowering dense-layer serialisation cycles and verification overhead.

Hardware results validate practical edge deployment: the FPGA implementation consumes 2,268 LUTs and 0.94 W while achieving 116 ms end-to-end latency, outperforming prior accelerators (QuantMAC, LPRE) and embedded platforms (Jetson Nano, Raspberry Pi). ASIC synthesis at 40nm reaches 1.56 GHz within a 3.29mm\textsuperscript{2} area and 1.65W power, confirming scalability beyond FPGA prototyping. Future work will explore runtime adaptive precision control and extension to multi-class acoustic scene recognition.

\begin{table}[!t]
\caption{Post-synthesis ASIC evaluation metrics at 40 nm for diverse AI accelerators}
\centering
\label{tab:asic_eval}
\resizebox{\linewidth}{!}{%
\begin{tabular}{|l|c|c|c|}
\hline
\textbf{Design} & \textbf{Freq. (GHz)} & \textbf{Area (mm\textsuperscript{2})} & \textbf{Power (W)} \\
\hline
JSSC'25 \cite{JSSC25} & 1.25 & 2.12 & 1.22 \\
TVLSI'25 \cite{MSDF-MAC} & 2.05 & 3.67 & 1.08 \\
TVLSI'25 \cite{Flex-PE} & 0.53 & 4.85 & 0.47 \\
ISCAS'25 \cite{Retro} & 1.93 & 4.73 & 5.71 \\
TCAS-I'22 \cite{ILM} & 1.46 & 10.80 & 1.02 \\
TRETS'23 \cite{GR-ACMTr} & 1.18 & 4.77 & 1.82 \\
\textbf{Proposed} & \textbf{1.56} & \textbf{3.29} & \textbf{1.65} \\
\hline
\end{tabular}}
\end{table}

\newpage

\bibliographystyle{ieeetr}
\bibliography{bib}

@ARTICLE{WST_CNN,
  author={Ali, Murtiza and Nathwani, Karan},
  journal={IEEE Signal Processing Letters}, 
  title={Exploiting Wavelet Scattering Transform and 1D-CNN for Unmanned Aerial Vehicle Detection}, 
  year={2024},
  volume={31},
  number={},
  pages={1790-1794}}

@ARTICLE{Flex-PE,
  author={Lokhande, Mukul and Raut, Gopal and Vishvakarma, Santosh Kumar},
  journal={IEEE Trans. {VLSI} Syst.}, 
  title={{Flex-PE: Flexible and SIMD Multiprecision Processing Element for AI Workloads}}, 
  year={2025},
  volume={33},
  number={6},
  pages={1610-1623}}

@inproceedings{publicsafety,
  author={M. Anwar and Z. Kaleem and A. Jamalipour},
  title={{ML-inspired sound-based amateur drone detection for public safety applications}},
  booktitle={IEEE Transactions on Vehicular Technology},
  volume={68},
  number={3},
  pages={2526--2534},
  year={2019}
}

@ARTICLE{9774975,
  author={McCoy, James and Rawal, Atul and Rawat, Danda B. and Sadler, Brian M.},
  journal={IEEE Trans. on Intelligent Transportation Syst.}, 
  title={Ensemble DL for Sustainable Multimodal UAV Classification}, 
  year={2023},
  volume={24},
  number={12},
  pages={15425-15434}}

@inproceedings{audio-based,
  author={Y. Wang and Z. Chu and I. Ku and E. C. Smith and E. T. Matson},
  title={A large-scale UAV audio dataset and audio-based UAV classification using CNN},
  booktitle={IEEE International Conference on Robot and Computing},
  pages={186--189},
  year={2022}
}

@article{environmental,
  author={S. Abdoli and P. Cardinal and A. Lameiras Koerich},
  title={End-to-end environmental sound classification using a 1D convolutional neural network},
  journal={Expert Systems with Applications},
  volume={136},
  pages={252--263},
  year={2019}
}

@ARTICLE{POLARON,
  author={Lokhande, Mukul and Jain, Akanksha and Vishvakarma, Santosh Kumar},
  journal={IEEE International Symposium on VLSI Design and Test}, 
  title={{Precision-aware On-device Learning and Adaptive Runtime-cONfigurable AI acceleration}}, 
  month=aug,
  year={2025}}

@article{Norrie2022,
  author={T. Norrie and N. Patil and D. H. Yoon},
  title={The Design Process for Google TPUv2 and TPUv3},
  journal={IEEE Micro},
  volume={41},
  pages={56--63},
  year={2022}
}

@ARTICLE{JSSC25,
  author={Li, Kai and Huang, Mingqiang and Li, Ang and others},
  journal={IEEE Journal of Solid-State Circuits}, 
  title={A 29.12-TOPS/W Vector Systolic Accelerator With NAS-Optimized DNNs in 28-nm CMOS}, 
  year={2025},
  volume={60},
  number={10},
  pages={3790-3801}}

@article{GR-ACMTr,
author = {Raut, Gopal and Karkun, Saurabh and Vishvakarma, Santosh Kumar},
title = {{An Empirical Approach to Enhance Performance for Scalable CORDIC-Based DNNs}},
year = {2023},
issue_date = {September 2023},
publisher = {Association for Computing Machinery},
address = {New York, NY, USA},
volume = {16},
number = {3},
issn = {1936-7406},
journal = {ACM Trans. Reconfigurable Technol. Syst.},
month = jun,
articleno = {39},
numpages = {32}
}

@article{GR-Neuro,
title = {{Data multiplexed and hardware reused architecture for DNN accelerator}},
journal = {Neurocomputing},
volume = {486},
pages = {147-159},
year = {2022},
issn = {0925-2312},
author = {Raut, Gopal and Biasizzo, Anton and Dhakad, Narendra and Gupta, Neha and Papa, Gregor and Vishvakarma, Santosh Kumar},
}

@INPROCEEDINGS{LPRE,
  author={Kokane, Omkar and Lokhande, Mukul and Raut, Gopal and Teman, Adam and Vishvakarma, Santosh Kumar},
  booktitle={IEEE International Symposium on Circuits and Systems}, 
  title={{LPRE: Logarithmic Posit-enabled Reconfigurable edge-AI Engine}}, 
  year={2025},
  volume={},
  number={},
  pages={1-5}}

@ARTICLE{ILM,
  author={Pilipović, Ratko and Bulić, Patricio and Lotrič, Uroš},
  journal={IEEE Transactions on Circuits and Systems I: Regular Papers}, 
  title={{A Two-Stage Operand Trimming Approximate Logarithmic Multiplier}}, 
  year={2021},
  volume={68},
month=jun,
  number={6},
  pages={2535-2545}}

@ARTICLE{MSDF-MAC,
  author={Cherati, Sahar Moradi and Barzegar, Mohsen and Sousa, Leonel},
  journal={IEEE Trans. on Very Large Scale Integr. Syst.}, 
  title={{MSDF-Based MAC for Energy-Efficient Neural Networks}}, 
  year={2025},
month=feb, 
  volume={},
  number={},
  pages={1-12}}

@ARTICLE{QuantMAC,
  author={Ashar, Neha and Raut, Gopal and others},
  journal={IEEE Access}, 
  title={{QuantMAC: Enhancing Hardware Performance in DNNs With Quantize Enabled Multiply-Accumulate Unit}}, 
  year={2024},
  volume={12},
  number={},
  pages={43600-43614}}

@ARTICLE{RAMAN,
  author={Krishna, Adithya and Rohit Nudurupati, Srikanth and others},
  journal={IEEE Internet of Things Journal}, 
  title={{RAMAN: A Reconfigurable and Sparse tinyML Accelerator for Inference on Edge}}, 
  year={2024},
  volume={11},
  number={14},
  pages={24831-24845}}

@ARTICLE{NULLHOP,
  author={Aimar, Alessandro and Mostafa, Hesham and others},
  journal={IEEE Trans. on NN and Learning Systems}, 
  title={NullHop: A Flexible CNN Accelerator Based on Sparse Representations of Feature Maps}, 
  year={2019},
  volume={30},
  number={3},
  pages={644-656}}

@ARTICLE{9857602,
  author={Lu, Jiahao and Liu, Dongsheng and Cheng, Xuan and others},
  journal={IEEE Trans. on Circ. and Syst. I}, 
  title={An Efficient Unstructured Sparse CNN Accelerator for Wearable ECG Classification Device}, 
  year={2022},
  volume={69},
  number={11},
  pages={4572-4582}}

@ARTICLE{Retro,
  author={Kokane, Omkar and Raut, Gopal and Ullah, Salim and others},
  journal={IEEE Computer Society Annual Symposium on VLSI (ISVLSI)}, 
  title={{Retrospective: A CORDIC-Based Configurable Activation Function for NN Applications}}, 
  year={2025},
  volume={},
  number={}}

@ARTICLE{ESL2,
  author={Mian, Feroz Ahmed and Zafar, Saima},
  journal={IEEE Embedded Systems Letters}, 
  title={{SoC-Based Implementation of 1-D CNN for 3-Channel ECG Arrhythmia Classification via HLS4ML}}, 
  year={2024},
  volume={16},
  number={4},
  pages={429-432}}

@INPROCEEDINGS{Iot1,
  author={Puduru, Vamsi Krishna and Pardhasaradhi, Bethi and Koorapati, Sagar and Cenkeramaddi, Linga Reddy},
  booktitle={2025 IEEE 14th International Conference on Communication Systems and Network Technologies (CSNT)}, 
  title={Lightweight Deep Learning Model for Airborne Object Classification}, 
  year={2025},
  volume={},
  number={},
  pages={230-234}}

@INPROCEEDINGS{Iot2,
  author={Raj, Goutham and Mathew, Raj I Susan and Cenkeramaddi, Linga Reddy},
  booktitle={2025 IEEE International Conference on Electronics, Computing and Communication Technologies (CONECCT)}, 
  title={Aerial Vehicle Classification Using a Custom Lightweight CNN: A Deep Learning Approach}, 
  year={2025},
  volume={},
  number={},
  pages={1-6}}

\end{document}